\documentclass{article}
\usepackage{arxiv}
\usepackage[utf8]{inputenc} % allow utf-8 input
\usepackage[T1]{fontenc}    % use 8-bit T1 fonts
\usepackage{hyperref}       % hyperlinks
\usepackage{url}            % simple URL typesetting
\usepackage{booktabs}       % professional-quality tables
\usepackage{amsfonts}       % blackboard math symbols
\usepackage{nicefrac}       % compact symbols for 1/2, etc.
\usepackage{microtype}      % microtypography
\usepackage{lipsum}
\usepackage{graphicx}
\usepackage{amsmath}
\usepackage{wrapfig}
\usepackage[sort&compress,comma,numbers]{natbib}

\title{Temporal Normalizing Flows}

\author{
  {Gert-Jan Both} \\
  CRI Research \\
  Universite Paris Descartes \\
  Paris, France \\
  \And
  Remy Kusters\thanks{Corresponding Author.} \\
  CRI Research \\
  Universite Paris Descartes \\
  Paris, France \\
  \texttt{remy.kusters@cri-paris.org}\\
}
\begin{document}

\maketitle

\begin{abstract}

Analyzing and interpreting time-dependent stochastic data requires accurate and robust density estimation. In this paper we extend the concept of normalizing flows to so-called “temporal Normalizing Flows” (tNFs) to estimate time dependent distributions, leveraging the full spatio-temporal information present in the dataset. Our approach is unsupervised, does not require an \textit{a-priori} characteristic scale and can accurately estimate multi-scale distributions of vastly different length scales. We illustrate tNFs on sparse datasets of Brownian and chemotactic walkers, showing that the inclusion of temporal information enhances density estimation. Finally, we speculate how tNFs can be applied to fit and discover the continuous PDE underlying a stochastic process.\\
\newline
Code and examples at github.com/PhIMaL/temporal\_normalizing\_flows

\end{abstract}

\section*{Introduction}

Density estimations from sparse time series data are ubiquitous to interpret probabilistic or stochastic phenomena in quantitative science, e.g. in econometrics \cite{zambom_review_2012}, variational inference \cite{rezende_variational_2015} and biological sciences \cite{manzo_review_2015}. In this latter application, single particle tracking (SPT) has become the method of choice to investigate the dynamics, structure and interaction of many molecules in a cellular context, allowing the observation of single molecule trafficking on the nanoscale throughout the cell. The obtained trajectories are typically interpreted as random walks and analyzed in terms of their mean squared displacement (MSD). This analysis provides insight into the underlying transport processes and has revealed its non-ergodicity and anomalous diffusive properties \cite{manzo_review_2015}. The main difficulty in SPT is linking the particles between the frames to create a trajectory; particles cross, thus exchanging their identity, or stop fluorescing completely \cite{manzo_review_2015}. Alternatively, there exists a rich mathematical literature studying trajectories in terms of walker densities \cite{klafter_first_2011}. This perspective provides an alternative way to extract transport properties from experimental data without the need to link the particles between the frames. Key to this approach is accurately inferring the evolving particle density, particularly when data is sparse. 

\begin{figure}[h]
    \centering
    \includegraphics[width=0.98 \linewidth]{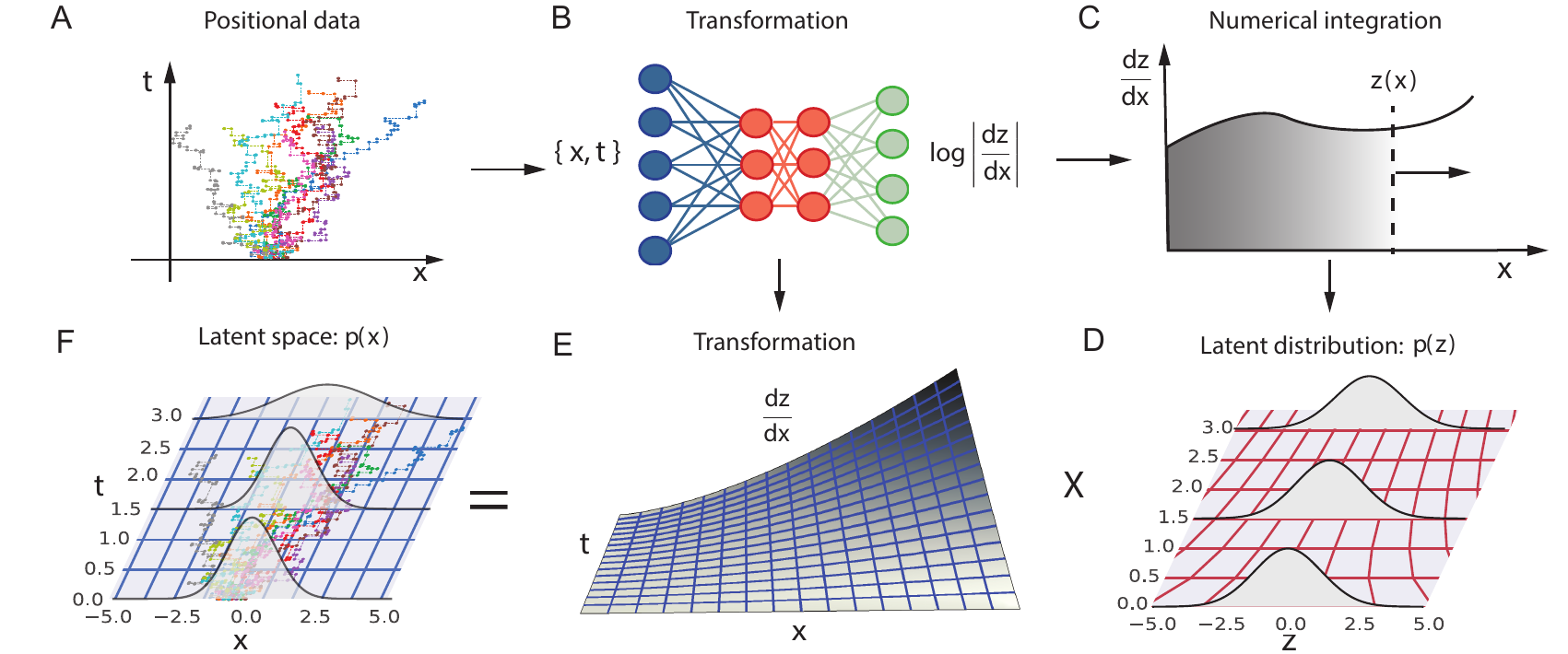}
    \caption{A schematic view of temporal Normalizing Flows (tNF). Multiple trajectories (\textbf{panel A}) are fed into a neural network, which outputs the log Jacobian of the mapping $z=f(x,t)$ (\textbf{panel B}). Next, the Jacobian is numerically integrated in space (\textbf{panel C}) to obtain the probability distribution in latent space (\textbf{panel D}). Multiplying this distribution with the Jacobian (\textbf{panel E}) provides an estimate of the walker density  (\textbf{panel F}). This process is iterated until the density in panel F reaches its maximum likelihood.}
    \label{fig:schematic_view}
\end{figure}{}

The classical approach to density estimation is binning. It provides an accurate density estimate when the sample-size is large, but becomes sensitive to the location of the bins when data becomes sparse. This method is also subject to a bias-variance trade-off; small-scale features are not captured when using oversized bins, whereas undersized bins will lead to a very noisy estimate. Alternatively, one can use continuous methods such as the Kernel Density Estimate (KDE). In KDE, a kernel is placed on each particle and the mean over all kernels then gives an estimate of the density. The resulting estimate is highly sensitive to the width of the kernel and although several automated estimators exist \cite{turlach_bandwidth_nodate, zambom_review_2012}, choosing the right width is a non-trivial task. While these techniques are firmly established, inferring the distribution of a random variable \emph{evolving through time} remains challenging. Explicitly including the temporal axis suppresses natural variations in the estimate by exploiting temporal correlations; samples taken closely together in time are likely to be only slightly different. To our knowledge, KDE and binning cannot include temporal dynamics under the constraint of particle conservation. In this paper we propose a novel technique based on normalizing flows, which is capable of handling these constraints. 

Normalizing Flows (NFs) learn an arbitrarily complex probability distribution by applying a series of transformations to a known distribution in a latent space. NFs originated in the field of machine learning and were initially applied to infer posterior distributions in the context of variational inference \cite{rezende_variational_2015}. They have been successfully applied as generative models (e.g. to generate novel faces \cite{kingma_glow:_2018}) and many papers showcase their capability as density estimators for 2D, time independent toy problems \cite{huang_neural_2018, chen_continuous-time_2017, dinh_density_2016}. NFs have several advantages as a density estimation technique: they are unsupervised, do not require an \textit{a-priori} length scale such as the bin- or kernel width and they can naturally accommodate several of such length scales across a dataset, a notoriously hard problem. Here we extend NFs to include temporal dynamics and hence name our approach \emph{temporal Normalizing Flows} (tNFs). 

The rest of the paper is organized as follows. In Section 2, we introduce and implement tNFs. Section 3 presents the application of tNFs on a multi-scale toy problem and datasets of Brownian and chemotactic particles. In Section 4 we present some further perspectives of this approach, in particular its potential as a physics informed density estimator and its ability to perform accurate density estimations on finite domains.
%The rest of the paper is organised as follows.
%\begin{itemize}
%    \item In section 2, we introduce and implement (temporal) normalizing flows. 
%    \item In section 3, we introduce our single particle models and show the results of the temporal-Neural flows.
%    \item In section 4, we showcase some initial results and applications of tNFs, and discuss future lines of research.
%\end{itemize}{}

\section*{Methods}
%We start this section by discussing normalizing flows, before deriving how to augment NFs to include time. As our implementation of tNFs differs significantly from other normalizing flow, we also discuss and motivate the details of our implementation. 

\paragraph{Normalizing Flows}
Consider a set of samples $\{x_0, x_1,..., x_n\}$ taken from an unknown distribution $p$. Given some model $p_\mathcal{X}(x; \theta)$, we can estimate $p$ by minimizing the negative log-likelihood of the model on the data,

\begin{equation}
    \mathcal{L} = -\sum_{i=1}^n \log p_\mathcal{X}(x_i; \theta).
    \label{eq:cost}
\end{equation}

To obtain an accurate estimate of $p$, the model $p_\mathcal{X}$ needs to be flexible enough. Normalizing flows \cite{chen_neural_2018, noe_boltzmann_2019, rezende_variational_2015} allow the construction of an \emph{arbitrary} model by applying an invertible transformation to a known probability density. Consider a random variable $z$ distributed by pdf $p_\mathcal{Z}$. Given an invertible transformation $x = g(z)$, $x$ is then distributed by $p_\mathcal{X}$, which is given by,

\begin{equation}
    \label{eq:CoV}
    p_\mathcal{X}(x) = p_\mathcal{Z}(z)\left| \frac{dg^{-1}(z)}{dx}\right|,\quad x = g(z) = f^{-1}(z).
\end{equation}{}

Typically, $\mathcal{X}$ is referred to as the real space and $\mathcal{Z}$ as the latent space and is usually a Gaussian. Normalizing flows learn the real to latent space mapping $z = f(x)$ and consequently the density $p_\mathcal{X}(x)$ by minimizing the negative log likelihood,

\begin{equation}
    \mathcal{L} = -\sum_{i=1}^n \log p_\mathcal{X}(x_i) = -\sum_{i=1}^n \log p_\mathcal{Z}(f(x_i)) + \log \left|\frac{df(x)}{dx} \right|_{x=x_i}.
    \label{eq:cost}
\end{equation}

\begin{wrapfigure}{r}{0.42\textwidth}
  \begin{center}
    \includegraphics[width=0.42\textwidth]{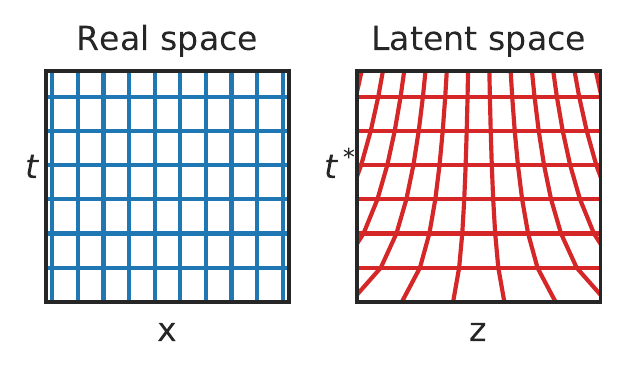}
  \end{center}
  \caption{Schematic interpretation of a temporal Normalizing flow. The regular grid in real space (left panel) gets deformed in latent space (right panel). The spatial coordinate gets stretched and compressed, but the temporal coordinate is equal to the real space time.}
    \label{fig:deformed_grid}
\end{wrapfigure}

\paragraph{temporal Normalizing Flows}
The normalizing flows as presented in the previous section cannot account for temporal dynamics. Nonetheless, our starting point for deriving the temporal NF is the N-dimensional equivalent of eq. \ref{eq:CoV}, 

\begin{equation}
    \label{eq:CoV-ND}
    p_\mathcal{X}(\mathbf{x}) = p_\mathcal{Z}(\mathbf{z})\left|\det J \right|,\quad \mathbf{z} = f(\mathbf{x}),
\end{equation}{}

where $J$ is Jacobian of $f(\mathbf{x})$. As we cannot write a conservation relation for the temporal axis, i.e. $\int p(x,t)dt \neq 1$, we cannot include it as an additional dimension in eq. \ref{eq:CoV-ND}, explaining why NFs cannot account for temporal dynamics. However, assume for now that such a construction is possible. The determinant of the Jacobian for a 1D temporally-varying distribution can then be written as, 

 \begin{equation}
    \label{eq:jacobian}
     \det J =\begin{vmatrix}
\frac{\strut\partial z}{\strut\partial x} & \frac{\strut\partial z}{\strut\partial t} \\
\frac{\strut\partial t^*}{\strut\partial x} & \frac{\strut\partial t^*}{\strut\partial t}\\
\end{vmatrix},
 \end{equation}{}
 
 where $z$ is the latent spatial coordinate and $t^*$ the latent temporal coordinate. Note that both are dependent on $x$ and $t$, i.e. $z=z(x,t)$ and $t^* = t^*(x, t)$. As the transformation of $t \to t^*$ is not allowed, \emph{the latent time $t^*$ must be equal to the real time $t$}, so that determinant of the Jacobian becomes,

\begin{equation}
    \label{eq:jacobian}
     \det J =\begin{vmatrix}
\frac{\strut \partial z}{\strut \partial x} & \frac{\strut \partial z}{\strut \partial t} \\
0 & 1
\end{vmatrix} 
=\frac{\strut \partial z(x, t)}{\strut \partial x}.
 \end{equation}{}
 
The 1D temporal Normalizing Flow can then be written as,

 \begin{equation}
    \label{eq:CoV-ND-t}
    p_{\mathcal{X}}(x, t) = p_\mathcal{Z}(z, t)\left|\frac{\partial z}{\partial x}\right|,\quad z = f(x, t).
\end{equation}{}

 We show a graphical interpretation of this in figure \ref{fig:deformed_grid}. While the temporal axis is not stretched or compressed, all frames are coupled through the mapping $z=f(x,t)$. Using a single mapping for the whole dataset prevents overfitting and suppresses natural variations in the estimate, as we will show in the results.

 \paragraph{Implementation}
 
The prime challenge of implementing NFs and tNFs in practice is finding a flexible yet invertible transformation. NFs are generally applied as generative models on high-dimensional data, requiring a computationally efficient method to evaluate the determinant of the Jacobian (see e.g. FFJORD \cite{grathwohl_ffjord:_2018}, Autoregressive flows 
\cite{papamakarios_masked_2017} or GLOW \cite{kingma_glow:_2018}). Spatio-temporal density estimation contains up to four dimensions, such that calculating the determinant of the Jacobian is not a computational constraint. This allows us to propose a relatively simple implementation. 

Wehenkel et al. \cite{wehenkel_unconstrained_2019} recently introduced a method for the construction of monotonic neural networks, independent of the networks' specific architecture. Building on the observation that a function is monotonic if its derivative is positive, they propose to constrain a neural network to positive outputs only and numerically integrate over the output to obtain a monotonic function. We slightly modify their approach and use an \emph{unconstrained} feed-forward neural network to model the log Jacobian, naturally leading to monotonic and hence invertible mapping $f(x,t)$. This leads to the following implementation for the tNF,
 
\begin{equation}
    \label{eq:implementation}
    p_{X}(x) = p_Z(z, t)e^{f(x, t)} ,\quad z = \int e^{f(x, t)}dx + z_0(t).
\end{equation}
Here $z_0(t)$ is a time dependent offset function. Both $f(x,t)$ and $z_0(t)$ are modeled by unconstrained neural networks with a tanh-activation function ($f(x,t)$ contains 3 hidden layers of 30 neurons and $z_0$ contains 1 hidden layer of 100 neurons). In the remainder of this work we choose a time independent Gaussian as latent distribution, $p_\mathcal{Z}$. We perform the integration in eq. \ref{eq:implementation} over a regular grid rather then integrating over the particles' positions. This approach scales with the size of the grid, rather than with the number of particles, works well when data is sparse and scales to higher dimensions.
 
\section*{Results}

We now demonstrate tNFs on three datasets: 
\begin{itemize}
    \item A \textbf{multi-scale toy problem} to show tNFs can accommodate different length scales in a single distribution;
    \item A dataset of \textbf{Brownian motion} to show how tNFs enhance density estimation for sparse datasets;
    \item A dataset of \textbf{chemotactic walkers} to show that tNFs 
    can correctly estimate a multi-modal, non-Gaussian density.
\end{itemize}

\paragraph{Multi-scale density estimation}

\begin{wrapfigure}{r}{0.45\textwidth}
  \begin{center}
    \includegraphics[width=0.45\textwidth]{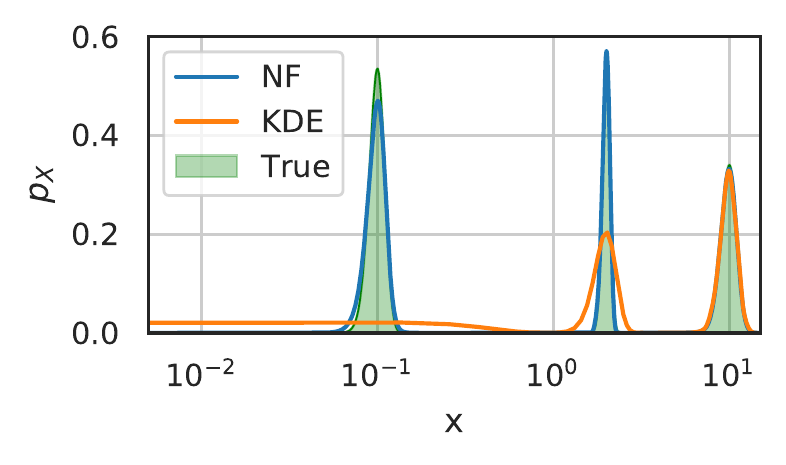}
  \end{center}
  \caption{Comparison of density estimation with different scales. $N=5000$ samples were taken from a density consisting of three Gaussians with widths $0.01$, $0.1$, and $1$, at respective locations $0.1, 2$ and $10$, and weights $0.013, 0.13$ and $0.85$. The KDE used a Gaussian kernel with the kernel width set by Scott's rule.} 
    \label{fig:multiscale}
\end{wrapfigure}

A key problem in density estimation is inferring an accurate distribution when vastly different length scales are present within a single dataset. Classical approaches such as binning and KDE require a single characteristic length scale, prohibiting an accurate estimate of a multi-scale distribution. We now show that normalizing flows, and by extension tNFs, are capable of accurately inferring such a distribution. 

We build an artificial distribution consisting of three normal distributions with standard deviations $\sigma$ of $0.01, 0.1$ and $1.0$ (thus spanning three orders of magnitude) and respective weights $0.013, 0.13$ and $0.85$. Figure \ref{fig:multiscale} shows the inferred distribution from 5000 samples for the NF and the KDE with Scott's rule determining the lengthscale. Observe that, as expected, the KDE is unable to accommodate the different scales and that due the different weighting of each peak, the widest is dominating the lengthscale estimation. Contrarily, the NF provide an accurate density estimate for all lengthscales present in the problem, independent of their weights.

%We stress that tNF do not require any prior information of the different lengthscales present in the data-set. 

\paragraph{Brownian motion}
Brownian motion is the most basic and ubiquitous random walk and thus an ideal test case to assess the performance of tNFs, comparing them to time independent NFs and classical binning. We generate a single trajectory for a Brownian random walker by the recursive relation, $ \vec{x}_{n+1} = \mathcal{N}(\vec{x}_n, \sqrt{2D\Delta t})$. Here $n$ is the step number with $x_0$ the initial position, $D$ the diffusive coefficient and $\Delta t$ the time step. In the limit of an infinite number of walkers, the \emph{walker density} $c$ is described by the diffusion equation, $\partial_t c = D\nabla^2 c$. 

\begin{figure}[h!]
    \centering
    \includegraphics[width=0.98 \linewidth]{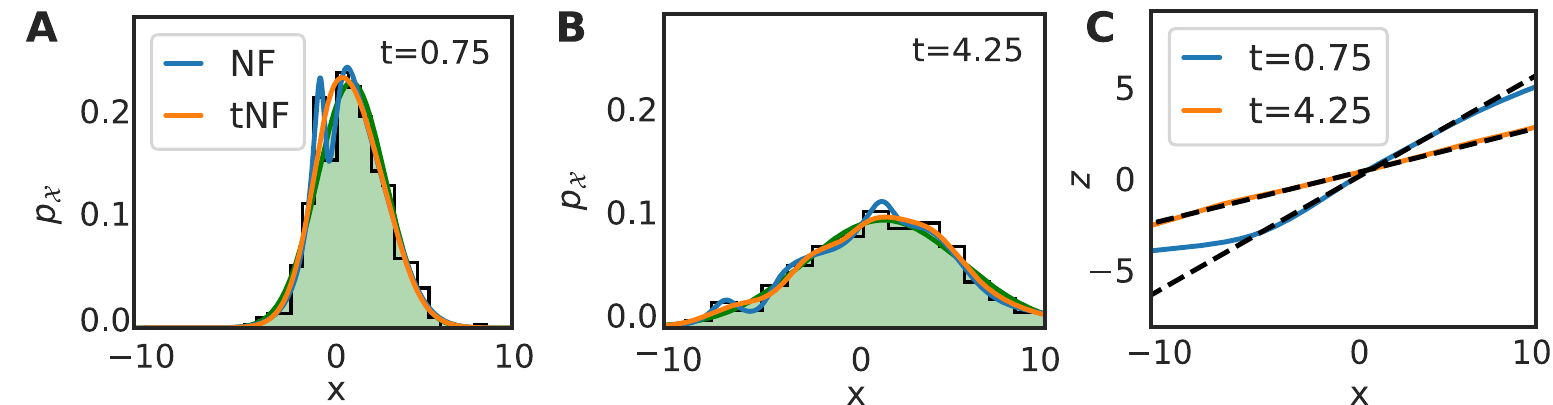}
    \caption{Results of inferring a Brownian walker density from $M=500$ walkers with $D=2$ and where a snapshot was taken every $\Delta t=0.1$ for $N=100$ frames. The initial position was sampled from a Gaussian placed at $x=1.5$ with width $\sigma=0.5$. Panel \textbf{a} and \textbf{b} compare the inferred density by binning, time independent normalizing flow (NF) and time dependent normalizing flow (tNF) at two different times. Panel \textbf{c} compares the learned mapping with the true mapping.}
    \label{fig:density_diffusion}
\end{figure}

Our dataset consists of $M=500$ walkers with $D=2.0$, with snapshots being taken every $\Delta t = 0.1$ for $N=100$ frames. The initial positions were sampled from a Gaussian centered at $x=1.5$ with width $\sigma=0.5$; in this case, the diffusion equation can be solved exactly and the solution behaves as a spreading Gaussian in time. We show the estimated density at $t = 0.75$ and $t = 4.25$ in figure \ref{fig:density_diffusion} (a) and (b) for the tNF, the time independent NF and binning. The tNF provides a significantly better density estimate than the time independent NF, illustrated by the difference in $\ell_2$ error; $2.6\cdot10^{-5}$ for the tNF and $7.4\cdot10^{-5}$ for the NF, averaged over frame 15 and 85. 

Normalizing flows are based on neural networks and hence prone to overfitting. We analyze the effect of overfitting in Appendix I and show that NFs overfit more strongly than tNFs and perform worse in terms of the $\ell_2$ error. We mainly attribute this improvement to the temporal correlations in the dataset, which suppresses the natural frame-to-frame variations in the density estimate. Nonetheless, tNFs are not immune to overfitting and we speculate performance could be enhanced by applying techniques such as early stopping.

For the diffusion equation the true mapping can be trivially derived. We compare it to the learned mapping in figure \ref{fig:density_diffusion}(c). It shows perfect agreement at $t=4.25$, but deviates from the true curve for $x<-5$ at $t=0.75$. As can be seen in figure \ref{fig:density_diffusion}(a), no samples were present in this domain, explaining the deviance. Nonetheless, it implies that the network does not generalize well outside the sampling domain. We speculate that techniques such as batch normalization or a different architecture for the network (a recurrent network, for example) might further improve performance. 

%Finally, figure \ref{fig:density_diffusion}d shows the distribution of the samples in latent space. As we fix the latent distribution to be Gaussian, if a proper mapping is learned the samples should indeed be distributed normally. As expected,the latent density (histogram) clearly follows the model (orange line) closely. 

\paragraph{Chemotaxis}
The Brownian motion presented in the previous section was a linear problem with a uni-modal, Gaussian solution. We now apply tNFs to so-called chemotactic walkers, a non-linear problem with a multi-modal solution. Bacteria and other micro-organisms sense gradients of chemicals throughout their environment and use this to guide their motion towards a food source. This effect is known as chemotaxis and is typically modelled by a random walker with a superimposed drift; $\vec{x}_{n+1} = \mathcal{N}(\vec{x}_n + \chi \nabla p(\vec{x}_n)dt, \sqrt{2Ddt})$, where $p$ is the chemical density and $\chi$ is the \textit{chemotactic sensitivity}, which controls the interaction between the chemical and the bacteria. In the infinite walker limit, the walker and chemical density are given by the Keller-Segel model: $ \partial_t c = \nabla \cdot (D_c \nabla c - \chi c\nabla p)$ and $\partial_t p = D_p \nabla^2 p - Kp$. Here $D_c$ and $D_p$ are the diffusion coefficients of the bacteria and the chemical respectively and a decay set by $K$ has been added to the chemical density. 

\begin{figure}
    \centering
    \includegraphics[width=0.98 \linewidth]{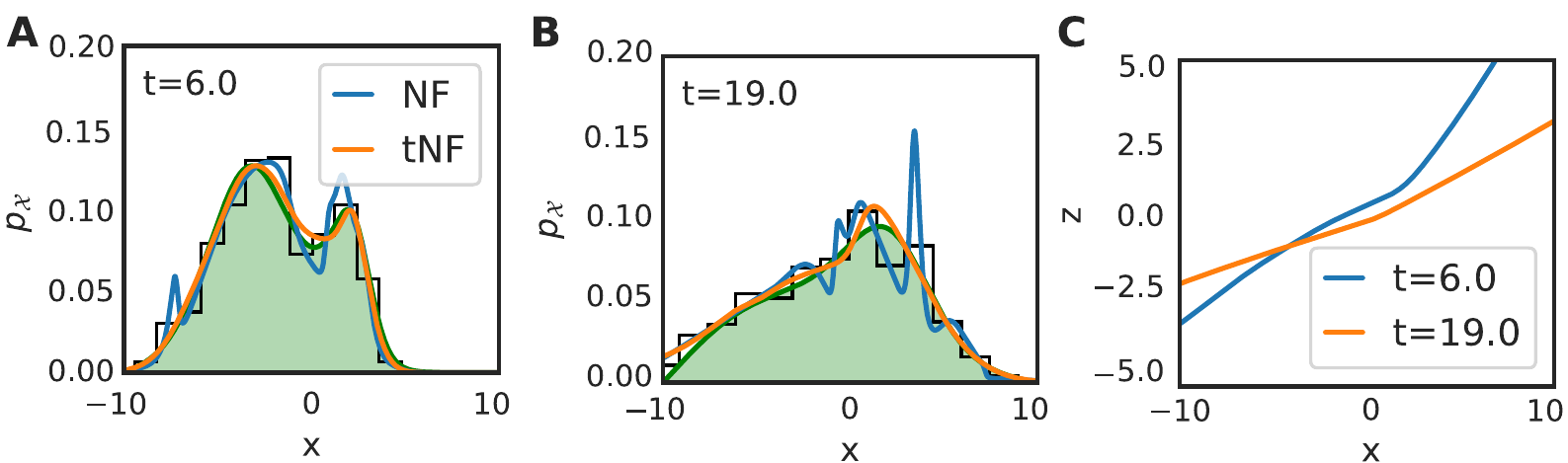}
    \caption{Results of inferring a chemotactic walker density from $M=500$ walkers with $D=0.5$ and chemotactic sensitivity $\chi = 10$, where a snapshot was taken every $\Delta t=0.1$ for $N=100$ frames. The initial position was sampled from a Gaussian placed at $x=-2.5$ with width $\sqrt{0.5}$, while the food source is a Gaussian located at $x=2.5$ with width $\sqrt{0.5}$, diffusing with $D=0.25$ and decaying at a rate of 0.05. Panel \textbf{a} and \textbf{b} compare the inferred density by binning, time independent normalizing flow (NF) and time dependent normalizing flow (tNF) at two different times. Panel \textbf{c} compares the learned mapping with the true mapping.}
    \label{fig:density_chemotaxis}
\end{figure}

Our dataset consisted of $M=500$ walkers with $D=0.5$ and we sampled the initial position from a Gaussian centred at $x=-2.5$. The food source was modelled by a Gaussian with diffusion coefficient $D=0.25$, centred at $x=2.5$; the walkers will thus drift towards food source over time. Figure \ref{fig:density_chemotaxis} shows a comparison of the time independent NF, tNF and the binning method. In figure \ref{fig:density_chemotaxis}(a) and (b) we find that the tNF leads to a significantly more accurate density estimation, illustrated by the difference in $\ell_2$ error ($1.45\cdot10^{-4}$ for the NF versus $1.8\cdot10^{-5}$ for the tNF, averaged over $t=6.0$ and $19.0$). The tNF captures the multi-modal distribution at $t=6.0$ excellently, without overfitting, contrarily to the time independent NF. The mapping, as shown in figure \ref{fig:density_chemotaxis}c, is non-linear, in contrast to the mapping obtained for the Brownian motion.

\section*{Perspective}

\begin{figure}
    \centering
    \includegraphics{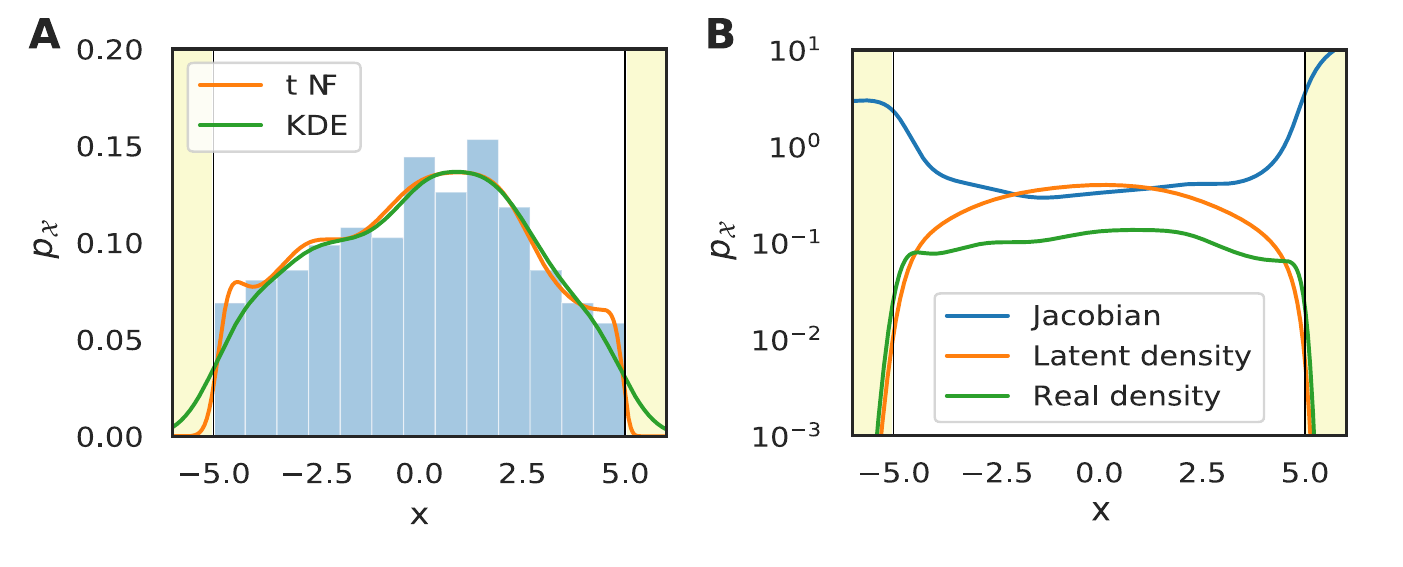}
    \caption{Results of inferring the Brownian walker density with two reflecting boundaries at $x=\pm 5$ with $M=1000$ walkers with $D=1.0$, where a snapshot was taken every $\Delta t=0.1$ for $N=50$ frames. The initial position was sampled from a Gaussian placed at $x=0.0$ with width $0.5$. Panel \textbf{a} compares the result at $t=4.5$ for the binning method, KDE and tNF. \textbf{b} deconstructs the result of the tNF by showing the Jacobian and latent density as a function of the real space coordinates.}
    \label{fig:boundary}
\end{figure}

\paragraph{Boundary conditions}
Density estimation near boundaries is often problematic \cite{botev_kernel_2010, malec_nonparametric_nodate}, as they introduce discontinuities in the profile. Applying KDE in such situations leads to non-zero probabilities past the boundary. We show here that NFs are less prone to these artifacts. In figure \ref{fig:boundary} we compare binning, KDE and tNF for 1000 random walkers between two reflective boundaries at $x=\{-5, 5\}$. We show the corresponding Jacobian and latent density in figure \ref{fig:boundary}b. At the boundaries, the latent density approaches zero, which must be compensated by the Jacobian to obtain the non-zero density of the true profile. How well the network is able to do this determines the quality of the estimate at the boundary and might lead to artifacts. To improve the density estimate near the boundary, we propose to use a latent distribution with finite support, e.g., the Epanechnikov kernel \cite{epanechnikov_non-parametric_1969}. However, this introduces a discontinuity in the cost function, leading to training issues. 

\paragraph{Physics Informed Normalizing Flows}

Physics Informed Neural Networks (PINNs)\cite{raissi_physics_2017} have emerged as a powerful yet simple method to include physical constraints in neural networks. They have been applied to (i) solve PDE's \cite{lu_deepxde:_2019}, (ii) infer parameters of a known equation \cite{raissi_inferring_2017} and (iii) perform model discovery \cite{both_deepmod:_2019}. Here, we propose Physics Informed Normalizing Flows (PINFs) to directly fit \emph{continuous models} to single particle data. Contrarily to PINNs, PINFs do not require an estimate of the density before fitting and explicitly conserve energy, mass or probability densities. By including the fitting in the cost function, PINFs form an end-to-end differentiable model to fit continuous models to discrete data. We construct it by adding the continuous model to the log-likelihood, analogously to a PINN,

\begin{equation}
\mathcal{L} = \sum_i \log p_X(x_i) + \frac{\lambda}{n}\sum_{i=0}^n\left|\partial_t p_X(x_i)-f(p_X, \partial_{x} p_X, ...) \right|^2.  
\label{eq:PINF}
\end{equation}

Here, $\lambda$ is a constant and sets the relative strength of the fitting term. The two terms in eq. \ref{eq:PINF} are of different origin (i.e. a likelihood term vs a MSE term) and hence are typically of different orders of magnitude. Consequently, training is more complex than PINNs, but preliminary testing on random walkers confirmed that PINFs are indeed capable of inferring the parameters of the PDE directly from the positional data. Further research however is required to improve the performance of these PINFs. 

%However, this required manual tuning of $\lambda$ and testing showed that the network did not reach the global minimum yet, leading to Further investigation is required to determine how to include the various cost functions and how to choose the magnitude of $\lambda$. While PINNs directly learn the differential equation, PINFs generate an estimate through the likelihood, leading to vastly different errors: PINNs are typically subject to independent, Gaussian noise, but in PINFs the error is not independent. A more robust regularization method (e.g. Huber loss) should hence be considered. 

%Finally, we discuss an issue we note as 'non-locality'. Since PINFs explicitly conserve a quantity, changing the density on one point propagates to other points. Although in PINNs this connection too exists through the weights of the neural network, it is much less strong. PINFs can thus not 'indepedently' adjust the result throughout its domain, and we speculate this makes training much harder. Nonetheless, we foresee a use for PINFs in applications involving energy and mass density, as one can be sure the solution conserves it. 

\paragraph{Discussion}
In this paper we have introduced temporal Normalizing Flows (tNFs), an extension of normalizing flows to estimate a time-varying probability density. We demonstrate that tNFs can naturally accommodate different length scales in a problem and outperform binning and time-independent normalizing flows, even when the density is non-Gaussian and multi-modal. tNFs use the full time series data to perform density estimation, rather than inferring the density one frame at a time. This exploits the temporal correlations in the data, which improves the performance of the neural network used to model the mapping. The use of an unconstrained monotonic neural network opens up the possibility of applying techniques such as batching and batch normalization, or even completely different architectures, e.g. RNNs.

We provide two perspectives, building on this work: (i) density estimation on a finite domain and (ii) discovering and fitting a PDE to the data. (i): Boundaries typically lead to discontinuous density profiles. In this situation, tNFs can provide a more accurate estimate of the true profile, compared to e.g. KDE. While the discontinuous density profile cannot be strictly modeled using a Gaussian latent distribution, we speculate that using a distribution with finite support could capture such discontinuities. (ii) Typically a continuous PDE can be derived for a time-dependent distribution. tNFs can be used to fit the corresponding PDE directly to positional data by simultaneously making an estimate of the density and fitting a PDE to the data. Rather than inferring parameters, we speculate that PINFs can also be used as PDE solvers (similar to \cite{lu_deepxde:_2019}) where energy or mass conservation is required. Our initial results with these physics informed normalizing flows are encouraging, but much work remains to be done, especially optimizing the training scheme. 

Our work fits in the wider context of temporal reasoning in machine learning. When applying generative modeling to a time series of images for example, the temporal axis must also be treated differently. Approaches based on modeling the latent time as a Gaussian process \cite{casale_gaussian_nodate} or as a Linear Gaussian State Space Model \cite{fraccaro_disentangled_nodate} have recently been been brought forward. We propose temporal normalizing flows could be used for similar time dependent applications. 

Access to the temporal dynamics of a process has both theoretical and practical benefits. Analysis and modeling experimental data is often limited to equilibrium processes \cite{noe_boltzmann_2019}, restricting the potential of the data at hand. Being able to study the temporal dynamics of such systems in terms of the underlying probability distribution or PDE opens up many opportunities in out-of-equilibrium science. We thus believe that tNFs can greatly aid the study of out-of-equilibrium processes.

\bibliographystyle{unsrtnat}  
\bibliography{references}  

\section*{Appendix I}
\begin{figure}
    \centering
    \includegraphics{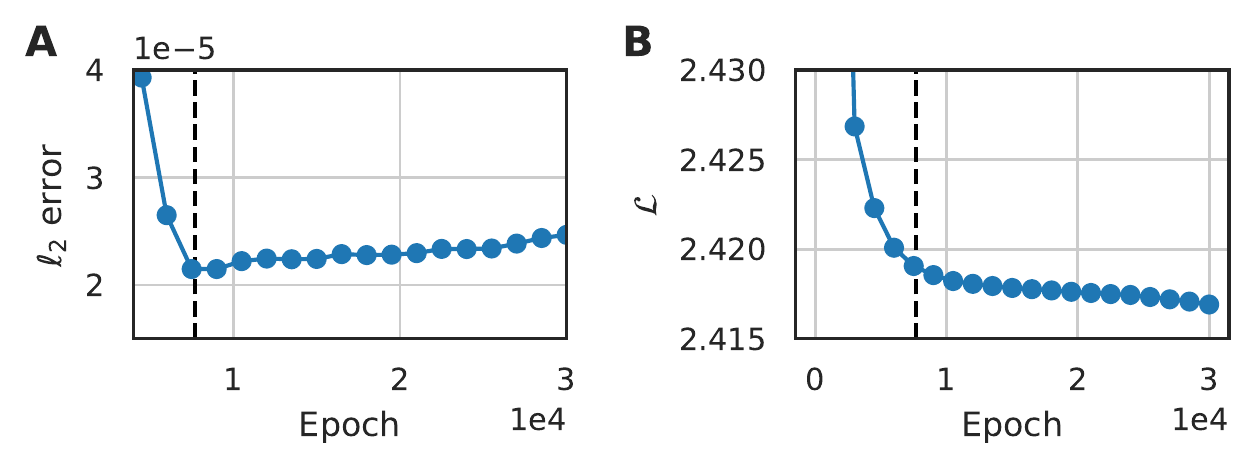}
    \caption{Performance of tNF during the training phase. Panel \textbf{a} shows the $\ell_2$ error with respect to the true solution as a function of the epoch for a Brownian walker dataset and panel \textbf{b} shows the corresponding negative log likelihood. The dashed black line indicates the location of the minimum $\ell_2$ error. We used a $M=500$ walkers with $D=2.0$ and took snapshots every $\Delta t=0.05$ for $N=100$ frames.}
    \label{fig:overfitting}
\end{figure}

In this appendix we study the effect of overfitting on the density estimation by comparing the $\ell_2$ error with the log-likelihood as a function of the training epoch in figure \ref{fig:overfitting}a and b. Here we performed a density estimate for 500 Brownian walkers with parameters identical as those selected in the main text. We delineate the minimum $\ell_2$ error with a black dashed line; note that this occurs after roughly 7000 epochs and that the error, with respect to the analytical solution increases upon training further. The negative log likelihood keeps decreasing however, corresponding to overfitting the solution. We found empirically that the minimum $\ell_2$ error occurs roughly at the elbow of the cost function, which for all cases considered is roughly at $\mathcal{O}(10^4)$ epochs so all the NF and tNF have been trained for 10000 epochs.

{}

\begin{figure}
    \centering
    \includegraphics{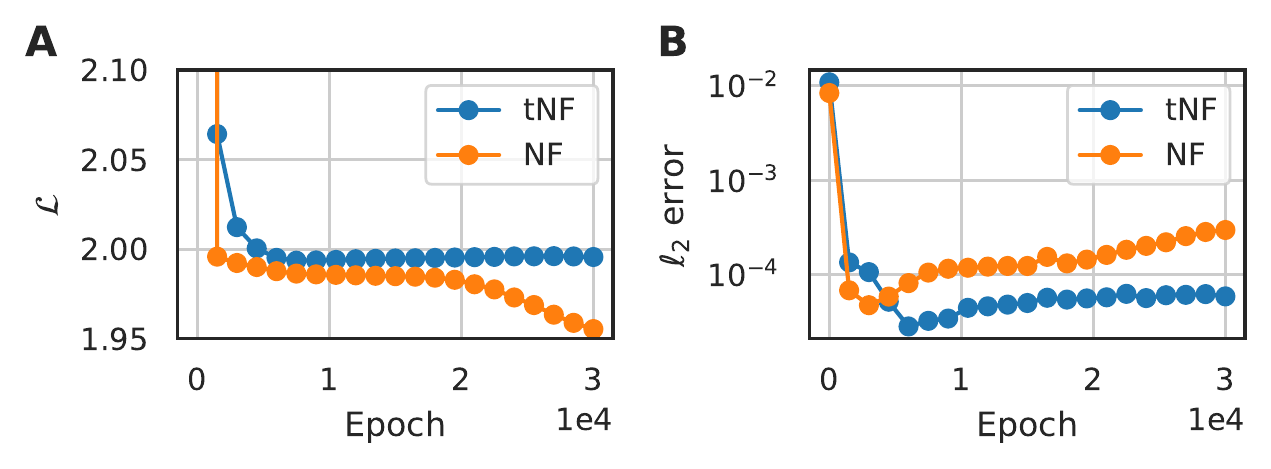}
    \caption{Comparison of performance of tNF with a time-independent NF during the training phase. Panel \textbf{a} shows the negative log likelihood for the NF and tNF at the representative frame $t=0.75$ as a function of the epoch and panel \textbf{b} shows the $\ell_2$ error with respect to the true solution for a Brownian walker dataset consisting of $M=500$ walkers with $D=2.0$ and took snapshots every $\Delta t=0.05$ for $N=100$ frames.}
    \label{fig:tnfvsnf}
\end{figure}

We show that tNFS are less prone to overfitting than NFs by comparing the log-likelihood and $\ell_2$ error for a single representative frame in figure \ref{fig:tnfvsnf} for both approaches. The log-likelihood of the time-independent NF keeps decreasing, leading to overfitting and an increased $\ell_2$ error. On the other hand, the tNF likelihood saturates and no longer decreases significantly after 10000 epochs, and neither does the $\ell_2$ error. 

{}

\end{document}